\documentclass[a4paper,11pt]{article}
\usepackage{pos}

\usepackage[symbol]{footmisc}
\usepackage[english]{babel}
\usepackage{graphicx}
\usepackage{graphics}
\usepackage{braket}
\usepackage{bbold}
\usepackage{amsmath}
\usepackage{nicefrac}
\usepackage{dcolumn}
\usepackage{bm}
\usepackage{slashed}
\usepackage{datetime}
\usepackage{mciteplus}
\usepackage{multirow}
\usepackage{bbold}
\usepackage{siunitx}
\usepackage{booktabs}
\usepackage{color, soul}
\usepackage[usenames,dvipsnames]{xcolor}
\usepackage{float}
\usepackage[utf8]{inputenc}
\usepackage[normalem]{ulem}
\usepackage{mathtools}
\usepackage{setspace}
\usepackage{comment}
\renewcommand{\thefootnote}{\fnsymbol{footnote}}

\newcommand{\drv}{{\rm d}}

\newcommand{\LQCD}{\Lambda_{\rm QCD}}

\newcommand{\Jps}{J/\psi}

\newcommand{\ecs}{\eta_c}
\newcommand{\ebs}{\eta_b}

\newcommand{{\HFNRevo}}{\tt HF-NRevo}

\title{Heavy-Flavor Fragmentation and Jet Structure \\ from HF-NRevo: Bridging to Heavy-Ion Collisions}
\ShortTitle{HF-NRevo: Bridging to Heavy-Ion Collisions}

\author*[a]{Francesco Giovanni Celiberto}
\author[a]{Francesca Lonigro}

\affiliation[a]{Universidad de Alcalá (UAH), Departamento de Física y Matemáticas, Campus Universitario, \\ Alcalá de Henares, E-28805, Madrid, Spain}

\emailAdd{francesco.celiberto@uah.es}
\emailAdd{francesca.lonigro@uah.es}

\abstract{We present recent progress on the Heavy-Flavor Non-Relativistic Evolution ({\HFNRevo}) framework, designed to describe leading-power fragmentation of heavy-flavored hadrons at moderate to large transverse momentum. 
Starting from NLO NRQCD calculations for all partonic channels into pseudoscalar quarkonia, we construct the {\tt NRFF1.0} collinear fragmentation functions via DGLAP evolution in a variable-flavor number scheme. 
We outline future prospects in the heavy-ion context, where {\HFNRevo} can serve as a baseline for modeling in-medium modifications of heavy-flavor fragmentation in nuclear collisions. 
Its accurate modeling of the partonic hierarchy and threshold effects makes it ideally suited to explore jet-quenching sensitivity, energy-loss mechanisms, and the emergence of medium-modified fragmentation functions in the quark-gluon plasma. 
Moreover, it provides a natural baseline for implementing in-medium hadronization scenarios, including quarkonium regeneration and fragmentation-function apparent-shape distortion.
These developments provide new handles for exploring heavy-flavor dynamics at the HL-LHC and future collider facilities.}

\FullConference{The European Physical Society Conference on High Energy Physics (EPS-HEP2025)\\
7-11 July 2025\\
Marseille, France\\}

\begin{document}
\maketitle

\section{Introductory remarks}
\label{sec:introduction}

In the study of fundamental interactions, hadrons containing open or hidden heavy flavors serve as key probes.
Heavy quarks, in particular, are central in searches for new physics due to their predicted couplings to beyond-Standard-Model particles.
Their masses, lying above the QCD confinement scale, also make them ideal for testing perturbative aspects of the strong force.
Quarkonia, often referred to as the “hydrogen atoms” of QCD~\cite{Pineda:2011dg}, offer a powerful window into the dynamics of hadron formation.
They bridge precision perturbative QCD with the exploration of proton structure.
For instance, hadronic decays of $S$-wave bottomonia allow precise extractions of $\alpha_s$~\cite{Brambilla:2007cz,Proceedings:2019pra}, while forward quarkonium production constrains the positivity of gluon PDFs at low~$x$ and low~$Q^2$~\cite{Altarelli:1998gn,Candido:2020yat}.
Quarkonia also play a central role in 3D imaging of the proton, both at small~\cite{Hentschinski:2020yfm,Celiberto:2018muu,Bolognino:2018rhb,Bolognino:2021niq,Celiberto:2019slj,Silvetti:2022hyc,Kang:2023doo} and moderate~$x$~\cite{Boer:2015pni,Lansberg:2017dzg,Bacchetta:2020vty,Bacchetta:2024fci,Celiberto:2021zww}.
Unresolved photoproduction of $\Jps$ plus a charm jet at the EIC is expected to probe the intrinsic-charm valence content of the proton~\cite{NNPDF:2023tyk,Flore:2020jau}.
Despite their importance, quarkonium hadronization remains a theoretical challenge.
Many models exist, yet none fully reproduce all experimental findings.
To address this, the effective field theory of Non-Relativistic QCD (NRQCD) was developed~\cite{Caswell:1985ui,Bodwin:1994jh}.
NRQCD assumes that physical quarkonia are superpositions of multiple Fock states, organized in a double expansion in $\alpha_s$ and the relative velocity $v$ of the $[Q \bar Q]$ pair.
Cross sections are written as sums of perturbatively calculable Short-Distance Coefficients (SDCs), each multiplied by a nonperturbative Long-Distance Matrix Element (LDME).
NRQCD provides a consistent framework to test quarkonium production mechanisms.
At low transverse momenta, the dominant process is the short-distance creation of a $[Q\bar Q]$ pair in the hard scattering.
At higher transverse momenta, however, single-parton fragmentation into quarkonia becomes increasingly relevant~\cite{Cacciari:1994dr}.

In this work, we study collinear fragmentation into pseudoscalar and vector quarkonia in the color-singlet channel, using a preliminary version of our {\tt NRFF1.0} fragmentation function (FF) sets~\cite{Celiberto:2025euy}.
These are built within a novel framework, Heavy-Flavor Non-Relativistic Evolution ({\HFNRevo})~\cite{Celiberto:2025euy,Celiberto:2024mex_article,Celiberto:2024bxu,Celiberto:2024rxa}, which employs NLO NRQCD inputs at the initial scale, evolves them via DGLAP, and includes a replica-based Monte Carlo treatment of missing higher-order uncertainties~\cite{Forte:2002fg}. 

\section{Quarkonium fragmentation within {\HFNRevo}}
\label{sec:HFNrevo}

\begin{figure*}[!t]
\centering

   \includegraphics[scale=0.38,clip]{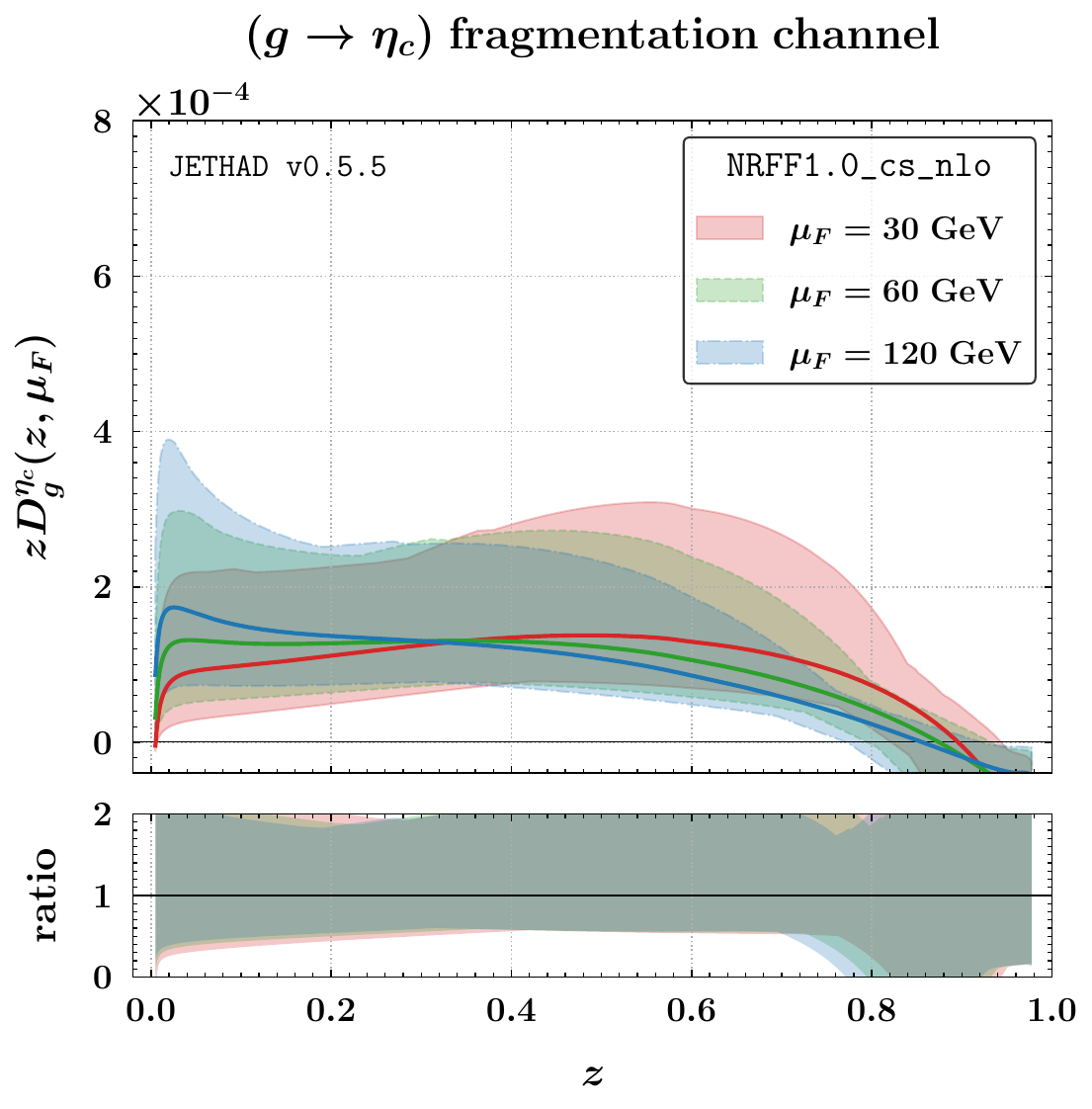}
   \hspace{0.15cm}
   \includegraphics[scale=0.38,clip]{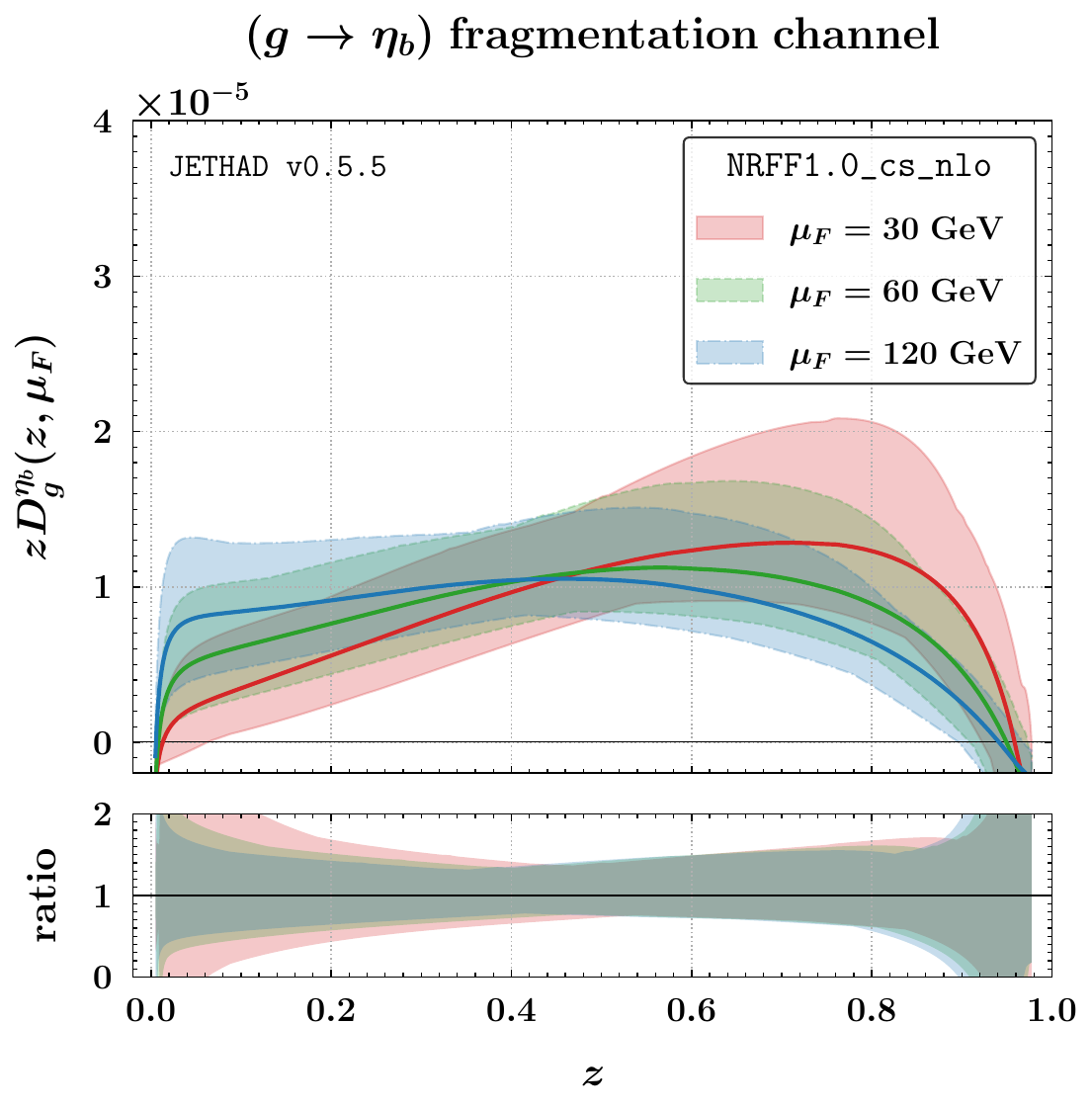}

\caption{NLO gluon to color-singlet $\ecs$ ($\ebs$) FFs. 
Plots adapted from~\protect\cite{Celiberto:2025euy}
}

\label{fig:FFs_bottom}
\end{figure*}

\begin{figure*}[!t]
\centering

   \includegraphics[scale=0.38,clip]{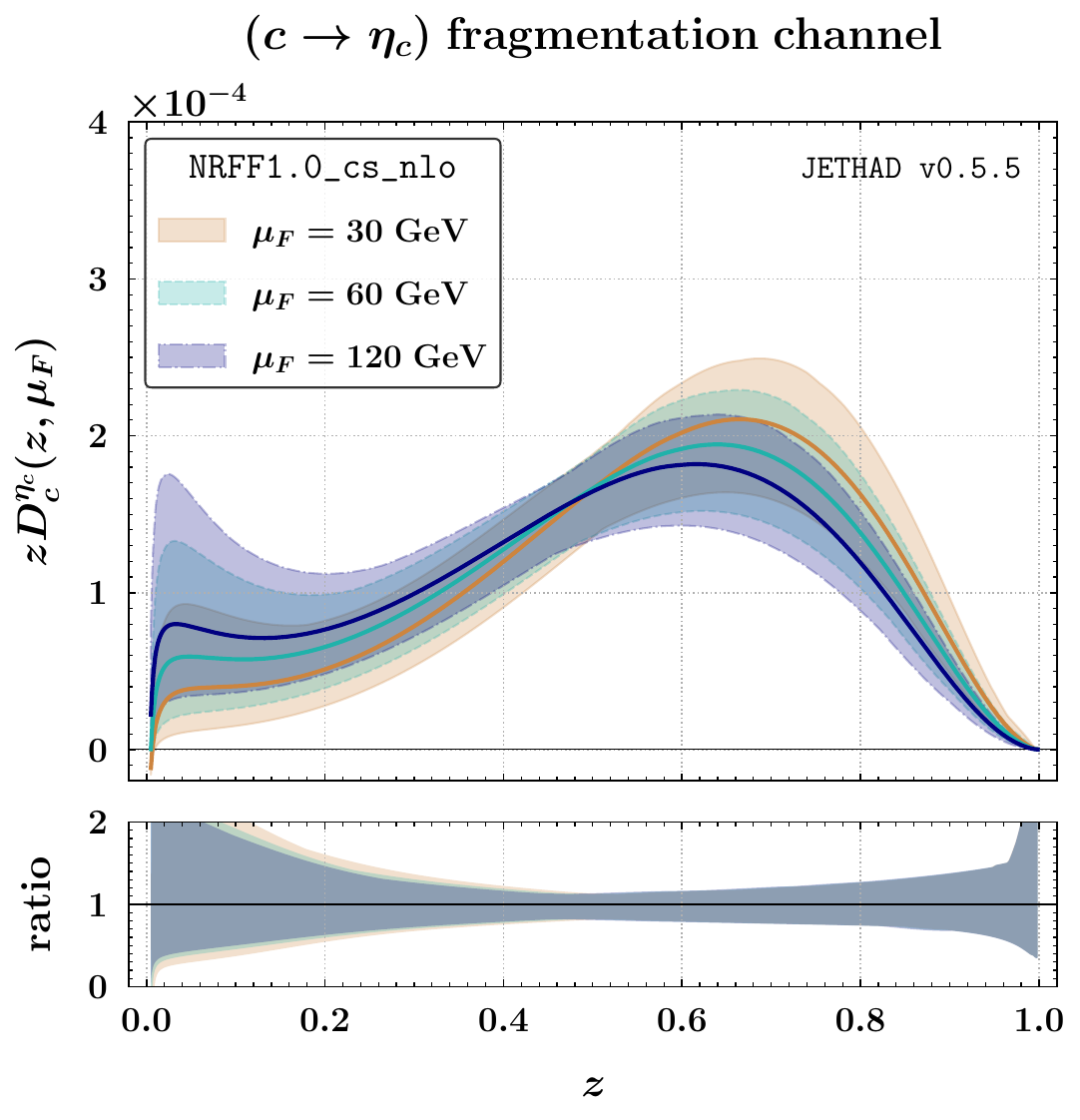}
   \hspace{0.25cm}
   \includegraphics[scale=0.38,clip]{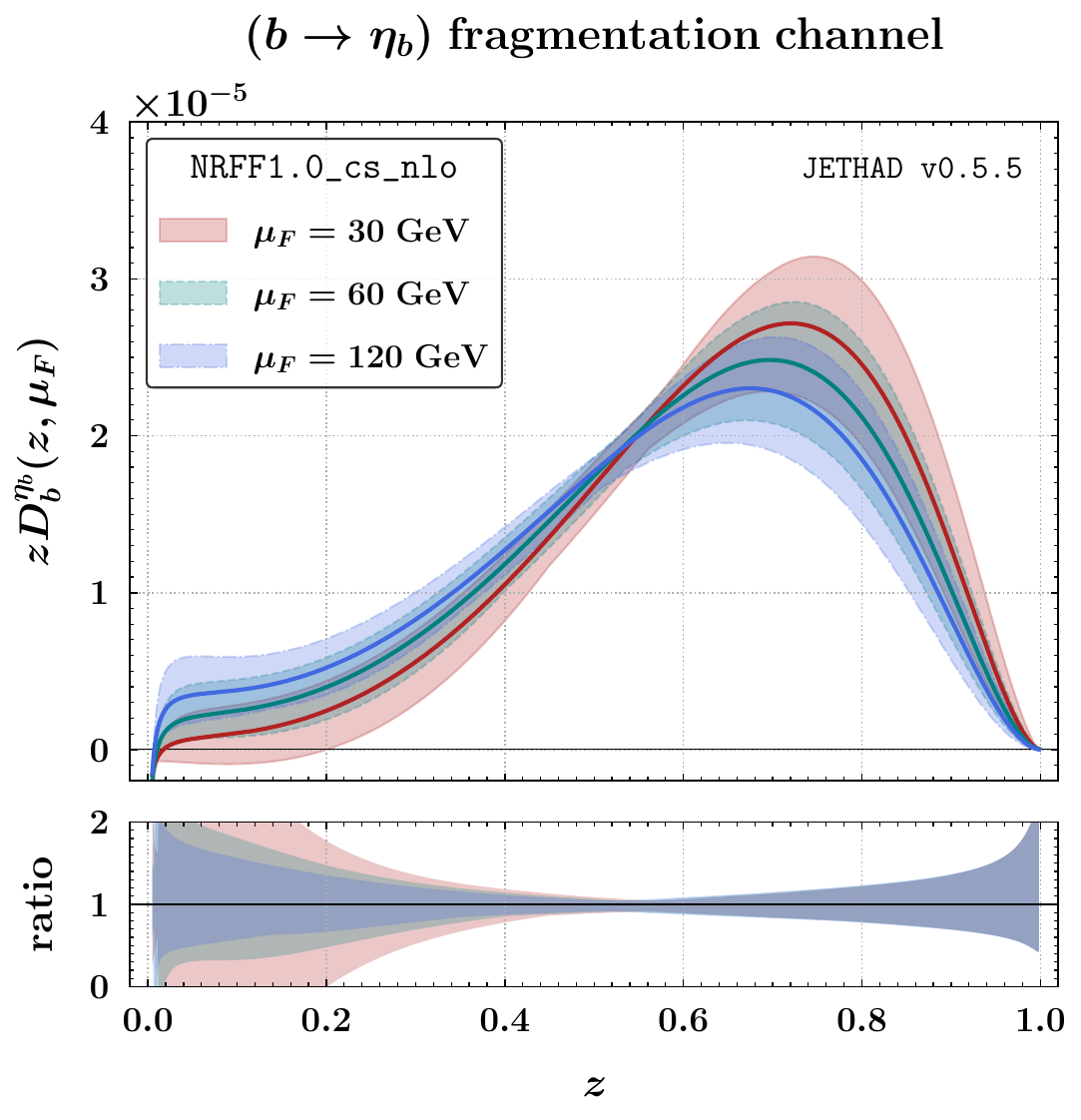}

\caption{NLO charm (bottom) to color-singlet $\ecs$ ($\ebs$) FFs. 
Plots adapted from~\protect\cite{Celiberto:2025euy}.
}

\label{fig:FFs_Q}
\end{figure*}

Given that the masses of constituent heavy quarks lie well above the QCD confinement scale, $\LQCD$, the initial conditions for FFs are expected to retain a perturbative character.
This motivates a fully consistent application of collinear factorization techniques.
To address this, we introduce a new methodology, dubbed {\HFNRevo}~\cite{Celiberto:2025euy,Celiberto:2024mex_article,Celiberto:2024bxu,Celiberto:2024rxa}, which provides a structured framework for defining, evolving, and estimating uncertainties in FFs for heavy quarkonia.
The {\HFNRevo} approach is anchored on three foundational pillars: interpretation, evolution, and uncertainty quantification.
First, in terms of physical interpretation, quarkonium production at low transverse momentum $|\vec{q}_T|$ can be understood as a two-parton fragmentation process within a Fixed-Flavor Number Scheme (FFNS).
This perspective naturally enables subsequent matching to a Variable-Flavor Number Scheme (VFNS) description.
Support for this interpretation comes from analyses of transverse-momentum-dependent (TMD) observables, where distinct singularity structures emerge in the high-$|\vec{q}_T|$ tails of shape functions~\cite{Echevarria:2019ynx} and collinear FFs at moderate $|\vec{q}_T|$~\cite{Boer:2023zit}.
The evolution of quarkonium FFs within {\HFNRevo} proceeds in two stages.
The first is a symbolic step, named {\tt EDevo}, which implements an expanded and decoupled DGLAP evolution to account for heavy-flavor thresholds.
This step is carried out using the symbolic engine of the {\tt JETHAD} framework~\cite{Celiberto:2020wpk,Celiberto:2022rfj,Celiberto:2022kxx,Celiberto:2020tmb,Bolognino:2021mrc,Celiberto:2021dzy,Celiberto:2021fdp,Celiberto:2022zdg,Celiberto:2022gji}.
The second step, {\tt AOevo}, consists of the full numerical all-order DGLAP evolution.
The final component of {\HFNRevo} concerns the systematic evaluation of missing higher-order uncertainties (MHOUs) arising from the treatment of evolution thresholds.
In particular, we adopt a scale-variation strategy that probes the sensitivity of FFs to changes in the renormalization and factorization scales entering the initial conditions.
These scales are varied simultaneously by a factor of two up and down from their central values.
This approach is conceptually aligned with recent developments in PDF uncertainty estimation, such as theory-covariance-matrix frameworks~\cite{NNPDF:2024dpb} and the {\tt MCscales} method~\cite{Kassabov:2022orn}.
To illustrate the behavior of quarkonium FFs in this framework, we focus on four representative fragmentation channels computed at NLO accuracy~\cite{Braaten:1993rw,Braaten:1993mp,Artoisenet:2014lpa,Zhang:2018mlo,Zheng:2021ylc,Zheng:2021mqr}, as recently implemented for pseudoscalar quarkonia $\ecs[^1S_0^{(1)}]$ and $\ebs[^1S_0^{(1)}]$ in Ref.~\cite{Celiberto:2025euy}.
Figure~\ref{fig:FFs_bottom} (left and right panels) shows the {\tt NRFF1.0} gluon-induced FFs for $(g \to \ecs)$ and $(g \to \ebs)$, evolved with $\mu_F$ spanning the same range as used in previous studies.
Similarly, Fig.~\ref{fig:FFs_Q} displays the charm-induced FFs for $(c \to \ecs)$ and $(c \to \ebs)$, where $\mu_F$ is varied from 30 to 120~GeV.

\section{Quarkonium-in-jet fragmentation}
\label{sec:onium_in_jet}

Jet substructure observables have recently emerged as powerful probes of the fundamental dynamics of the strong interaction.
In addition to deepening our understanding of QCD, they offer promising avenues for discovering signatures of physics beyond the Standard Model.
A detailed investigation of the internal structure of jets (particularly those containing heavy-flavored hadrons) provides critical insight into perturbative and nonperturbative aspects of QCD~\cite{Procura:2009vm,Bauer:2013bza,Chien:2015ctp,Maltoni:2016ays,Kang:2017glf,Metodiev:2018ftz,Marzani:2019hun,Kasieczka:2020nyd,Nachman:2022emq,Dhani:2024gtx}.
Among the most relevant observables are those sensitive to the detection of a specific hadron within the jet.
From the perspective of collinear factorization, this process is described by the formalism of Semi-Inclusive Fragmenting Jet Functions (SIFJFs).
At leading power, the SIFJF for a parton $i$ fragmenting into an identified quarkonium state ${\cal H}_Q$ inside a jet takes the form~\cite{Kang:2017yde}
\begin{equation}
 \label{eq:SIFJF}
 {\cal F}_i^{\cal H}(z, z_{\cal H}, \mu_F, {\cal R}_{\cal J}) \, = \,
 \sum_{j=q,\bar{q},g} \int_{z_{\cal H}}^1 \frac{\drv \zeta}{\zeta} \,
 {\cal S}(z, z_{\cal H}/\zeta, \mu_F, {\cal R}_{\cal J}) \,
 D_j^{\cal H}(\zeta, \mu_F) \;,
\end{equation}
with $D_j^{\cal H}(\zeta, \mu_F)$ denoting the standard $[j \to {\cal H}_Q]$ FF channel, and ${\cal S}(z, z_{\cal H}/\zeta, \mu_F, {\cal R}_{\cal J})$ standing for the perturbative fragmenting jet coefficients~\cite{Baumgart:2014upa}, known at NLO for anti-$\kappa_T$ and cone jet algorithms~\cite{Kang:2016ehg}.
In Eq.~\ref{eq:SIFJF}, the variable $z$ denotes the ratio between the light-cone momentum of the outgoing jet and that of the initiating parton $i$.
Similarly, $z_{\cal H}$ represents the ratio between the light-cone momentum of the hadron identified inside the jet and that of the jet itself.
The parameter ${\cal R}_{\cal J}$ specifies the jet radius.

\section{Bridging to heavy ions}
\label{sec:heavy_ions}

In heavy-ion collisions, the production of quarkonium states is strongly affected by the presence of a deconfined QCD medium. 
One observes a significant suppression of charmonia and bottomonia relative to the proton–proton baseline. 
This suppression originates from two main mechanisms~\cite{Vogt:1999cu,Satz:2000bn}. 
The first is \emph{color screening}, which reduces the binding potential between the heavy quark and antiquark due to the presence of thermal color charges in the quark-gluon plasma (QGP)~\cite{Shuryak:1980tp,Heinz:2000bk,Braun-Munzinger:2015hba}. 
As a result, loosely bound states dissociate more easily, leading to a sequential melting pattern that reflects their binding energies.
For example, ground states such as $J/\psi$ and $\Upsilon$ are more robust than their first excited counterparts, $\psi(2S)$ and $\Upsilon(2S)$, as well as the pseudoscalar partners $\eta_{c}$ and $\eta_{b}$~\cite{Matsui:1986dk,Grandchamp:2003uw,GayDucati:2003xa}.
The second mechanism is gluon-induced \emph{dissociation}, where thermal gluons scatter inelastically with quarkonia, driving their breakup. 
At higher collision energies, this process is partially balanced by quarkonium \emph{regeneration} via the recombination of independently produced heavy quark pairs in the QGP. 
The resulting yield is governed by the interplay between suppression and regeneration, whose relative contributions depend on temperature, lifetime, and the density of heavy quarks~\cite{Andronic:2008gu,Grandchamp:2003uw}.
To capture these effects quantitatively, transport models and hydrodynamic simulations are commonly employed, with lattice QCD inputs for in-medium binding energies and dissociation rates. 
However, uncertainties remain large, especially at high transverse momentum and for pseudoscalar states such as $\eta_c$ and $\eta_b$.

In this context, the {\HFNRevo} framework offers a promising vacuum baseline to define and compute medium-modified FFs for quarkonia. 
Due to its scale-resolved evolution and threshold-sensitive structure, {\HFNRevo} can be extended to model parton energy loss, quenching sensitivity, and the emergence of medium-modified FFs in nuclear environments. 
It also provides a natural starting point for implementing in-medium hadronization scenarios, such as statistical recombination and color de-excitation.
Moreover, {\HFNRevo} allows one to formulate observables sensitive to the deformation of fragmentation patterns in the QGP. 
In particular, the concept of fragmentation-function apparent-shape distortion (FF-ASD) can be used to parametrize deviations from vacuum FFs due to thermal broadening, dissociative reshaping, or delayed formation time. These observables can be exploited using high-$p_T$ quarkonium-tagged jets at the HL-LHC and future colliders, providing new tomographic probes of the QGP.

\section{Conclusions and Outlook}
\label{sec:conclusions}

We have introduced the {\HFNRevo} framework as a new methodology for constructing quarkonium FFs in the collinear limit.
Based on this scheme, we developed the first release of the {\tt NRFF1.0} set~\cite{Celiberto:2025euy}, featuring color-singlet initial-scale inputs for all parton channels, computed in NRQCD at NLO.
DGLAP evolution is performed with a consistent threshold prescription, and uncertainties are estimated through a Monte Carlo replica-like analysis accounting for missing higher-order effects.
The {\tt NRFF1.0} functions~\cite{Celiberto:2025euy} are intended to supersede earlier sets such as {\tt ZCW19}$^+$ and {\tt ZCFW22}, used in recent studies of vector quarkonia~\cite{Celiberto:2022dyf,Celiberto:2023fzz} and $B_c$ mesons~\cite{Celiberto:2022keu,Celiberto:2024omj}.
Thanks to their theoretical consistency, they are well-suited for experimental programs at the HL-LHC~\cite{Chapon:2020heu,Amoroso:2022eow}, the EIC~\cite{AbdulKhalek:2021gbh,Khalek:2022bzd,Abir:2023fpo}, and future facilities~\cite{AlexanderAryshev:2022pkx}, as well as for benchmarking artificial-intelligence-based FF extractions~\cite{Allaire:2023fgp,Hammou:2023heg,Costantini:2024xae}.
Future developments include the incorporation of color-octet channels for vector states~\cite{Cho:1995vh,Cacciari:1996dg}, implementation of a general-mass VFNS~\cite{Cacciari:1998it,Forte:2010ta,Guzzi:2011ew}, and applications to rare~\cite{Celiberto:2025ogy} and exotic hadrons~\cite{Celiberto:2023rzw,Celiberto:2024mrq,Celiberto:2024mab,Celiberto:2024beg,Celiberto:2025dfe,Celiberto:2025ziy,Celiberto:2025ipt}.
A longer-term goal is the extension of {\HFNRevo} to quarkonium-in-jet fragmentation, enabling the study of jet substructure via quarkonium-modulated angularities and resummation-sensitive observables.
In parallel, {\HFNRevo} offers a promising baseline for modeling in-medium modifications of heavy-flavor fragmentation in nuclear collisions.
Its accurate treatment of partonic thresholds and flavor hierarchies makes it ideal for exploring jet quenching, energy loss, and the emergence of medium-modified FFs in the quark-gluon plasma.
It also enables realistic in-medium hadronization scenarios, including quarkonium regeneration and FF apparent-shape distortion, offering new probes of heavy-flavor dynamics at the HL-LHC and beyond.

\section*{Acknowledgments}
\label{sec:acknowledgments}

We are supported by the Atracci\'on de Talento Grant n. 2022-T1/TIC-24176 (Madrid, Spain).

\vspace{-0.05cm}
\begingroup
\setstretch{0.6}
\bibliographystyle{bibstyle}
\bibliography{biblography}
\endgroup

\end{document}